\documentclass[12pt]{article}
\usepackage{graphicx}
\include{dspace12}
\def\bra{\langle}
\def\ket{\rangle}

\begin{document}

\begin{flushright}
hep-ph/0510298
\end{flushright}

\begin{center}
{\Large\bf Bilarge neutrino mixing from supersymmetry
with high-scale nonrenormalizable interactions}\\[20mm]

\renewcommand{\thefootnote}{\fnsymbol{footnote}}
Biswarup Mukhopadhyaya$^{1,}$\footnote[1]{Electronic address:
biswarup@mri.ernet.in}, 
Probir Roy$^{2,}$\footnote[2]{Electronic address:
probir@theory.tifr.res.in} 
and Raghavendra Srikanth$^{1,}$\footnote[3]{Electronic address:
srikanth@mri.ernet.in}\\
$^1${\em Harish-Chandra
Research Institute,\\
Chhatnag Road, Jhusi, Allahabad - 211 019, INDIA}\\
$^2${\em Department of Theoretical Physics, Tata Institute of Fundamental 
Research,\\
Homi Bhabha Road, Mumbai - 400 005, INDIA}\\[20 mm]

\end{center}

\begin{abstract}

We suggest a supersymmetric (SUSY) explanation of neutrino masses and 
mixing, where nonrenormalizable interactions in the hidden sector generate 
lepton number violating Majorana mass terms for both right-chiral 
sneutrinos and 
neutrinos. It is found necessary to start with a superpotential including 
an array of gauge singlet chiral superfields. This leads to nondiagonal 
$\Delta L = 2$ mass terms 
and almost diagonal SUSY breaking $A$-terms. As a result, the observed 
pattern of bilarge mixing can be naturally explained by the simultaneous 
existence of the seesaw mechanism and radiatively induced masses. 
Allowed ranges of parameters in the gauge singlet 
sector are delineated, corresponding to each of the cases of normal hierarchy,
inverted hierarchy and degenerate neutrinos.

\end{abstract}

{\bf PACS indices: 12.60.Jv, 14.60.Lm, 14.60.Pq, 14.60.St}
\newpage

\section{Introduction}

With the evidence in favor of neutrino masses and mixing at a convincing level
now, attempts to seek the role of physics beyond the 
standard model in the neutrino sector 
are acquiring enhanced degrees of urgency. As it is, the lack 
of naturalness of the mass of the Higgs boson in the standard model is a 
strong pointer towards new physics around the TeV scale. Since tiny masses 
can be explained by appealing to energy scales much higher than the 
electroweak scale (for example in the 
seesaw mechanism), it is appropriate to link neutrino mass generation 
to new physics options at the TeV scale or above.

Supersymmetry (SUSY) is a frequently explored possibility for new TeV-scale 
physics. Its capability for solving the naturalness problem being an
accepted fact, serious efforts are on at accelerators 
to see signals of SUSY, broken with an intra-supermultiplet mass splitting 
$\cal O$(TeV). Does SUSY play a role 
in providing the requisite new
physics component in the masses and mixing pattern of neutrinos? This is the
question that we would like to address. Often one has to go beyond the 
minimal SUSY standard model (MSSM) in order to find satisfactory
mechanisms which can achieve this end. Though a fair amount of work has been 
done in this area \cite{Mukhopadhyaya}, one is yet to have a satisfactory 
answer to the following central question related to the neutrino sector. Why is
the mixing pattern of neutrinos, with two large and one small mixing angles, 
so drastically different from that in the quark sector, where the mixing 
between the three generations can be cryptically described by the progression 
- small, smaller, smallest?

In this paper we approach the above problem with the idea that the difference
between the two mixing patterns
arises from some aspects of the SUSY model which are specific to neutrinos and
with no counterparts in the quark sector. For this, we make use of
nonrenormalizable terms arising from high-scale physics.
Such terms, coupling some hidden sector (gauge singlet) chiral 
superfields 
to the MSSM ones, are suppressed by the Planck mass $M_P$ or some
power of it. If these
terms violate lepton number , they can lead to 
Majorana masses for 
neutrinos. When, in addition, there are superfields containing 
right-chiral 
neutrinos, contributions to the neutrino mass matrix can come 
not only from the well-known 
seesaw mechanism but also radiatively via one-loop diagrams containing right 
chiral sneutrinos. Though both these 
contributions have been included in earlier works \cite{Arkani1,Arkani2,
Borzumati}, a clear explanation
of the different character of mixing for neutrinos vis-a-vis quarks
has been lacking without the imposition of some additional restriction on the 
low-energy theory. The Froggatt-Nielsen mechanism 
\cite{FNmech} is an example of such additional theoretical inputs. 
The literature, of course, is rich with uses of various other 
symmetries \cite{Symmetry}, as well as of `anarchy' in the neutrino mass 
matrix \cite{Anarchy}. Drawing inspiration from all these approaches, 
we suggest an alternative justification of 
bilarge neutrino mixing by postulating
an array of gauge singlet chiral superfields with flavor-dependent 
nonrenormalizable couplings to neutrinos. The specific superpotential that
yields the desired results is formulated, and consistency with
the observed suppression of flavour-changing neutral currents (FCNC) in 
the lepton sector is used as a constraint. One further has 
radiative contributions, pertaining only to the neutrino mass matrix. 
These are due to the fact that 
the right chiral sneutrinos may acquire gauge singlet $\Delta L =2$ mass
terms. The scenario for neutrinos then immediately becomes quite 
distinct from that in the quark sector.

Using the standard seesaw as well as the above-mentioned radiative 
contributions, we have examined
whether high-scale parameters, such as the vacuum expectation values 
(VEV) of scalar as well as of auxiliary components of the gauge singlet chiral
superfields, crucial to this mechanism, are in otherwise acceptable ranges 
of values. For instance, the Higgsino mass ($\mu$) parameter needs to be 
around the weak scale for the desired implementation of the spontaneous
breakdown of electroweak symmetry.
Such analyses are carried out for the three alternative possibilitiess of the 
neutrino mass spectrum allowed by the neutrino oscillation data,
namely, normal hierarchy, inverted hierarchy and
degenerate neutrinos. The simultaneous importance of the radiative as well as 
the seesaw contributions enables us to acquire in all the three scenarios 
substantial regions (of different extent in each case) in the parameter space 
of our model that correspond to acceptable solutions.

In section 2 we describe the (by now well-known) structure of the 
neutrino mass matrix $M_\nu$ for bilarge mixing. The SUSY model is constructed
and the elements of $M_{\nu}$ are consistently generated in section 3; we 
also show at the end of this section how FCNC processes, induced at one loop,
are suppressed. The SUSY parameter space, answering to each of the
specific scenarios of normal hierarchy, inverted hierarchy and degenerate 
neutrinos, is analyzed in section 4. We comment on some related possibilities  
in section 5. Section 6 contains our summary and conclusions.

\section{Facts about neutrinos}

There is experimental evidence now that neutrinos have tiny masses. We shall 
work within the scheme of three light active neutrinos, not including the
possibility of an additional light sterile one suggested by the LSND data till
results from the ongoing mini-Boone experiment settle the issue. We shall 
further assume CPT conservation.
While there is an upper bound \cite{Cosmo} of $\sim 1$ eV on the sum of 
neutrino mass eigenvlaues from cosmology, the lack of observation of 
neutrinoless double beta decay implies an upper bound of 
$\sim 0.3$ eV on the absolute value of the 
the {\it 11}-element of the neutrino mass 
matrix \cite{Beta}. On the other hand, the accumulating data from solar,
atmospheric, accelerator and reactor neutrino experiments persistently point 
\cite{Strumia} towards neutrino oscillations. These data identify favored 
regions of small but distinct mass-squared
separation of the three different physical neutrino states. At the same time, 
they also indicate that the mixing between the second and the third 
families is near maximal,
that between the first and the second is 
large, while the one between the first and the third families 
is restricted to a small angle. 
In perfect analogy with the Cabibbo-Kobayashi-Maskawa (CKM) matrix in the 
quark sector, the
three-flavour neutrino mixing matrix can be 
parameterized as 
\begin{eqnarray}
U = \left( \begin{array}{ccc}
        c_{12}c_{13} & s_{12}c_{13} & s_{13}e^{-i\delta } \\
        -s_{12}c_{23}-c_{12}s_{23}s_{13}e^{i\delta } &
        c_{12}c_{23}-s_{12}s_{23}s_{13}e^{i\delta } & s_{23}c_{13} \\
        s_{12}s_{23}-c_{12}c_{23}s_{13}e^{i\delta } &
        -c_{12}s_{23}-s_{12}c_{23}s_{13}e^{i\delta } & c_{23}c_{13}
     \end{array} \right).
\end{eqnarray}
In Eq. (1) $c_{ij} = \cos\theta_{ij}$, $s_{ij} = \sin\theta_{ij}$, $i,j$ being
family indices which run from 1 to 3 (Majorana phases have been neglected 
here). We work in the basis where the charged 
lepton mass matrix is diagonalized. While solar (reactor) neutrino 
(antineutrino) studies suggest that 
$\theta_{12}\simeq 32^o$ \cite{MSW, Solar}, 
the atmospheric neutrino deficit needs $\theta_{23}$ to be 
$\sim 45^o$ \cite{atmos} and data from reactors require that 
$\theta_{13}\leq 13^o$ \cite{Chooz}. Thus a pattern of bilarge mixing emerges.

The above pattern allows one to construct a candidate neutrino mass matrix
in terms of the mass eigenvalues $m_1, m_2, m_3$. To start with, let us take 
$\theta_{23} = \frac{\pi}{4}$ and $\theta_{13} = 0$ approximately, keeping 
$\theta_{12}$ free to be large. Then the corresponding transformation matrix 
can be written as 
\begin{eqnarray}
U_\nu = \left( \begin{array}{ccc}
        c & s & 0 \\
        -\frac{s}{\sqrt{2}} & \frac{c}{\sqrt{2}} & \frac{1}{\sqrt{2}} \\
        \frac{s}{\sqrt{2}} & -\frac{c}{\sqrt{2}} & \frac{1}{\sqrt{2}}
        \end{array} \right),
\end{eqnarray}
where $s=\sin\theta_{12}$ and $c=\cos\theta_{12}$. The neutrino Majorana mass 
matrix in the flavor basis can now be obtained by transforming the 
diagonal matrix with the above:
\begin{eqnarray}
M_\nu &=& U_\nu \left( \begin{array}{ccc}
        m_1 & & \\
         & m_2 & \\
         & & m_3
        \end{array} \right) U^T_\nu
                \nonumber \\
      &=& \left( \begin{array}{ccc}
        m_1c^2+m_2s^2 & \frac{cs}{\sqrt{2}}(-m_1+m_2) &
        \frac{cs}{\sqrt{2}}(m_1-m_2) \\
        \frac{cs}{\sqrt{2}}(-m_1+m_2) & \frac{1}{2}(m_1s^2+m_2c^2+m_3) &
        \frac{1}{2}(-m_1s^2-m_2c^2+m_3) \\
        \frac{cs}{\sqrt{2}}(m_1-m_2) & \frac{1}{2}(-m_1s^2-m_2c^2+m_3) &
        \frac{1}{2}(m_1s^2+m_2c^2+m_3)
        \end{array} \right).
\end{eqnarray}
Thus we see that the requirement of bilarge mixing commits one to a particular
structure of the mass matrix where, of course, the relative magnitudes of the 
entries depend on the eigenvalues. For more precise information one has to
take up the specific scenario of normal/inverted hierarchy or that of
degenerate neutrinos. In our study, we attempt to link the diagonal 
and off-diagonal mass terms of $M_\nu$ to the parameters
of the SUSY model at high scale and see what the different scenarios tell 
us about the model parameters themselves.

\section{The SUSY model and neutrino masses}

\subsection{Required features of the model} 

The model that we adopt is motivated by a number of recent 
works~\cite{Arkani1,Arkani2,Borzumati}. In ~\cite{Arkani2}, for example, a 
minimal extension of the MSSM, including a right-handed neutrino, is used. 
There the terms of the effective Lagrangian responsible for neutrino masses 
are 
\begin{eqnarray}
 {\cal L}_{eff} = \frac{1}{M_P}\left([X^\dagger NN]_D +
                \left[XLNH_u\right]_F\right) + {\rm h.c.},
\end{eqnarray}
where $M_P$ is Planck scale and coupling coefficients of order unity have been
suppressed. In Eq. (4), the chiral field $X$ can acquire both 
SUSY violating and SUSY conserving VEVs~\cite{Yanagida}. The above terms 
can be responsible for seesaw masses of order ${\frac{ {F^2_X} }{M_P^3}}$ 
for the neutrinos (recalling the need to have 
$F_X \sim \bra x\ket^2 \sim M_P
\bra H_u\ket$, to ensure the generation of other superparticle masses in 
the TeV range)
. In addition, there can be radiative contributions to the 
mass matrix from $\Delta L = 2$ sneutrino mass terms after SUSY breaking. 
These can arise with the help of a term 
$\frac{1}{M_P^3}[X^\dagger XX^\dagger NN]_D$, 
making contributions that can dominate over seesaw  masses in certain regions 
of the parameter space. 

We aim to explain the bilarge mixing pattern of neutrinos by extending 
such a model. As mentioned earlier, the basic philosophy is to envision some 
feature of neutrinos, which has no counterpart in the quark sector, as being 
responsible for the observed bilarge mixing. There are two features of this 
kind in such a model: (a) the right-chiral 
neutrino sector and (b) the corresponding right-chiral sneutrino sector,
with provisions of $\Delta L = 2$ terms in each. In our
approach, each of these sectors is attributed with a $3\times 3$ mass matrix 
structure which plays a crucial role in the contribution to the radiative as 
well as seesaw mass terms. 
Moreover, we postulate an array of gauge singlet chiral superfields $X_{ij}$.
Following these propositions, Eq.(4) is generalized to 
\begin{eqnarray}
 {\cal L}_{eff} = \frac{1}{M_P}\left([X_{ij}^\dagger N^iN^j]_D +
                \left[X_{ij}L^iN^jH_u\right]_F\right) + {\rm h.c.},
\end{eqnarray}
where there is a summation over flavor (i.e. generation) indices $i,j$. The 
choice of the above Lagrangian can be motivated by a global symmetry 
$G_F\times G$ \cite{Arkani1}. 
The factor $G$  helps in solving
the $\mu$-problem, and keeps the spontaneous SUSY breaking scale 
$\sqrt{F_X}$ low enough for the superparticle spectrum in the observable 
sector to be around TeV energies. The summation over family indices in 
Eq. (5) can be justified by the global symmetry  $G_F$. This essentially 
means that the hidden sector chiral superfields $X_{ij}$ interact with those 
of the visible sector with such a global symmetry and that the 
low-energy flavour structure is the artifact of such interactions. 
In case the $X_{ij}$'s acquire  SUSY violating VEVs, then different soft SUSY
breaking terms will arise from the 
nonrenormalizable interactions shown in Eq. (5) and other higher dimension
terms compatible with all symmetries of the theory. These 
arise in addition to 
the soft terms that have analogues in the squark sector.

Schematic expressions can be written for the neutrino 
and sneutrino mass terms, thus obtained, and for the soft SUSY breaking 
A-terms as well as for the corresponding terms in 
the SUSY Lagrangian leading to them. They are obtainable from the following 
realizations.

\begin{equation}
{\frac{1}{M_P}} {\int X_{ij}L^iN^jH_u {d^2}\theta} \longrightarrow 
({m_D})_{ij} \simeq  
 \langle x_{ij}\rangle M_{EW}/{M_P}\\
\end{equation}

\begin{equation}
{\frac{1}{M_P}} {\int X_{ij}^{\dagger} N^{i}N^{j} {d^4}\theta} 
\longrightarrow ({m_R})_{ij} \simeq  2 F_{X_{ij}}^*/{M_P}\\
\end{equation}

\begin{equation}
{\frac{1}{M_P}}{\int X_{ij}L^iN^jH_u {d^2}\theta} \longrightarrow 
M_{EW}{A_{ij}}
\simeq {F_X}_{ij} M_{EW}/{M_P}\\
\end{equation}

\begin{eqnarray}
{\frac{1}{3!M_P^3}}{\int [X_{ik}^{\dagger} X_{kl} X_{lj}^{\dagger} + X_{ik} 
X_{kl}^{\dagger} X_{lj}^{\dagger} + X_{ik}^{\dagger} X_{kl}^{\dagger} X_{lj} 
N^i N^j] {d^4}\theta} \longrightarrow 
	\\ \nonumber
(m^2_N)_{ij} \simeq  \frac{-1}{6M_P^3}
[\bra F^*_{X_{ik}}\ket \bra F_{X_{kl}}\ket \bra x^*_{lj}\ket + 
 \bra F^*_{X_{lj}}\ket \bra F_{X_{kl}}\ket \bra x^*_{ik}\ket + 
\bra F^*_{X_{kl}}\ket \bra F_{X_{ik}}\ket \bra x^*_{lj}\ket + 
	\\ \nonumber
	 \bra F^*_{X_{lj}}\ket \bra F_{X{ik}}\ket 
\bra x^*_{kl}\ket + \bra F^*_{X{ik}}\ket \bra F_{X{lj}}\ket \bra x^*_{kl}\ket 
+ \bra F^*_{X_{kl}}\ket \bra F_{X_{lj}}\ket \bra x^*_{ik}\ket],
\end{eqnarray}

\noindent
where $M_{EW} = v/\sqrt 2 = 174$ Gev. The above expressions
are up to unknown multiplicative factors occurring in the SUSY Lagrangian.  
It is also assumed that the Dirac mass matrix, generated by canonical Yukawa 
couplings (as in the quark sector) arising from renormalizable terms in the 
superpotential, has very small off-diagonal elements. Also, in addition to the
$\Delta L = 2$ mass terms shown above, there may be L-conserving mass terms 
for right-chiral sneutrinos as a result of soft SUSY breaking.

Nondiagonal A-terms can potentially contribute to FCNC processes such as 
$\mu\to e\gamma$ and hence need to be suppressed. Therefore, we wish to have a
structure where $F_{X_{ij}}$ vanishes for $i \neq j$. On the other hand, 
contributions to off-diagonal terms in the neutrino mass matrix are 
essential for bilarge mixing. Such terms will be made to arise from seesaw as 
well as radiative processes. As we shall show below, both of these are driven 
by the VEVs of the scalar components of the X-superfields. We must therefore 
have nonzero $\langle x_{ij} \rangle$ for $i \neq j$. 

Thus we require nonrenormalizable terms in the superpotential involving
the chiral superfields $X_{ij}$, with a rather interesting
complementarity between the diagonal and nondiagonal elements of the
array. The diagonal ones can have nonvanishing F-term VEVs and thus can
generate a diagonal A-matrix, whereas the off-diagonal elements must
have vanishing F-term VEVs, though the corresponding scalar VEVs
must be nonvanishing. A superpotential, in which the above characteristics can
be achieved, is presented below.

\subsection{The superpotential}

It was noticed in the previous subsection that bosonic components of the array
of chiral superfields $X_{ij}$ should 
acquire SUSY violating and SUSY conserving VEVs in a complementary 
manner to be able to generate the required neutrino masses and mixing pattern. 
In order to achieve this, we first demonstrate a simple situation in 
which the auxiliary and the scalar components of a single chiral superfield 
acquire SUSY violating and SUSY conserving VEVs respectively. Thereafter we 
generalize this to an array of such superfields $X_{ij}$. 

Consider a set of hidden sector fields, for which the superpotential is of the 
form \cite{Arkani1,Arkani2} 
\begin{eqnarray}  
W = S(Y\bar{Y} - \mu^2_1) + Y^2\bar{X}^{\prime } + \bar{Y}^2X
        \nonumber \\
    +S'(X\bar{X} - \mu^2_2) + X^2\bar{Z} + \bar{X}^2Z.
\end{eqnarray}
Here the $R$-charge for the chiral field $X$ is $\frac{1}{3}$, while the  
$R$-charges of the remaining chiral fields can be chosen so that $W$ has 
$R$-charge 2. 
The explicit assignment of $R$-charges will be shown 
after generalizing this to an array  $X_{ij}$. 
The superpotential $W$, shown above, leads to a scalar potential $V$ 
which has local minima. The position of the abosolute minimum  
depends on the parameters $\mu_1$ and $\mu_2$.

Case(1): If $|\mu_1|<|\mu_2|$, the true minimum occurs at $\bra y\ket 
= \bra\bar{y}\ket = \bra s\ket = \bra s^\prime\ket = \bra\bar{x}^\prime\ket 
= \bra z\ket = \bra\bar{z}\ket = 0$ and $|\bra x\ket| = |\bra\bar{x}\ket| = 
\frac{|\mu_2|}{\sqrt{3}}$. In this case we have $\bra x\ket\neq 0$ but 
$F_X = 0$. Thus there is a nonzero scalar VEV, but SUSY breaking does 
not show up in the obserbable sector.\\

Case(2): If $|\mu_2|<|\mu_1|$, the true minimum occurs at $|\bra y\ket| 
= |\bra\bar{y}\ket| = \frac{|\mu_1|}{\sqrt{3}}$ and $\bra x\ket = 
\bra\bar{x}\ket = \bra s\ket = \bra s^\prime\ket = \bra\bar{x}^\prime\ket 
= \bra z\ket = \bra\bar{z}\ket = 0$. In this case $\bra x\ket = 0$ but 
$F_X\neq 0$ and SUSY is spontaneously broken.\\

Utilizing these lessons, we make a straightforward
genralization of the above superpotential to include an array of chiral 
superfields $X_{ij}$:

\begin{eqnarray}
W = \sum_{i,j=1}^{3} W_{ij},
\end{eqnarray}
where
\begin{eqnarray}
W_{ij} = S_{ij} (Y_{ij}\bar{Y}_{ij} - \mu^2_{ij}) +
         Y^2_{ij}\bar{X}_{ij}^{\prime } + \bar{Y}_{ij}^2X_{ij} +
        \nonumber \\
        S_{ij}^\prime (X_{ij}\bar{X}_{ij} - \mu^{\prime 2}_{ij}) +
         X_{ij}^2\bar{Z}_{ij} + \bar{X}_{ij}^2Z_{ij}.
\end{eqnarray}
The $R$-charges of various superfields in this scheme are shown in Table 1.\\

\begin{table}
\begin{center}
\begin{tabular}{|c|c|ccccccccc|}  \hline

Hidden sector & Field & $X_{ij}$ & $\bar{X}_{ij}$ & $\bar{X}_{ij}^{\prime}$ & 
$Y_{ij}$ & $\bar{Y}_{ij}$ & $S_{ij}$ & $S_{ij}^\prime$ & $Z_{ij}$ & 
$\bar{Z}_{ij}$ 
  \\ \cline{2-2} 
 & R-charge & $\frac{1}{3}$ & $-\frac{1}{3}$ & $\frac{11}{3}$ & $-\frac{5}{6}$ 
& $\frac{5}{6}$ & 2 & 2 & $\frac{8}{3}$ & $\frac{4}{3}$ \\  \hline
Visible sector & Field & $Q^i$ & $L^i$ & $U^i$ & $D^i$ & $E^i$ & $N^i$ & 
$H_u$ & $H_d$  &   \\ \cline{2-2} 
 & R-charge & $\frac{1}{2}$ & $\frac{1}{2}$ & $\frac{1}{2}$ & $\frac{1}{2}$ & 
$\frac{1}{2}$ & $\frac{1}{6}$ & 1 & 1 & \\ \hline

\end{tabular}
\end{center}
\caption{R-charges of hidden and visible sector superfields.}
\end{table}

With the above R-charge assignments, the 
nonrenormalizable interactions relevant to us are all R-invariant.
Moreover, the $\mu$-parameter is obtained in the desired range from the 
term ${\frac{1}{M_P}}\sum \int (S_{ij}^\dagger + S^{\prime\dagger}_{ij}) 
H_u H_d d^{4}\theta$, yielding 

\begin{equation}
\mu \simeq {\frac{F_S}{M_P}}, 
\end{equation}

\noindent
where $F_S$ includes a sum over indices.
Thus this scenario has the additional virtue of explaining 
the value of $\mu$ around the electroweak scale,
so long as the hidden sector F-terms can be justified to be at an intermediate
scale $\sim 10^{11}$ GeV.

The minima of the scalar potential arising from the above superpotential 
occur at $\bra s_{ij}\ket = \bra s_{ij}^\prime\ket = 0$, 
$\bra\bar{x}_{ij}^\prime\ket = 0$, $\bra z_{ij}\ket = \bra\bar{z}_{ij}\ket 
= 0$ and further depend on the parameters $\mu_{ij},\mu^\prime_{ij}$. We 
choose these parameters in the following way:
\begin{enumerate}
        \item For $i = j$, choose $|\mu_{ij}^\prime|  < |\mu_{ij}|$ so that
                $F_{X_{ij}} \neq 0$ and $\bra x_{ij}\ket = 0$.
        \item For $i \neq j$, choose $|\mu_{ij}| < |\mu_{ij}^\prime|$ so that
                $F_{X_{ij}} = 0$ and $\bra x_{ij}\ket \neq 0$.
\end{enumerate}
The generation of off-diagonal entries in the neutrino 
Majorana mass matrix
via nondiagonal $\langle x_{ij}\rangle$ is ensured by this
potential. On the other hand, we have secured a diagonal form for 
$F_{X_{ij}}$, thus suppressing contributions to FCNC processes  
from A-terms. This interesting complementarity is achieved rather
naturally by postulating in the superpotential the presence of some mass 
parameters $\mu_{ij}$ and $\mu^\prime_{ij}$ and their relative hierarchies. 
Though these parameters may all be broadly of the same 
order, the relative magnitudes of the primed and unprimed ones can 
naturally be quite different for different members of the array.   
The actual suppression of FCNC processes will be demonstrated in further 
detail in a later subsection.

\subsection{Neutrino mass matrix}

Schematic expressions for neutrino and 
sneutrino mass
terms, induced in this scenario, have already been shown. Now we obtain 
the exact entries in the neutrino
mass matrix. These will enable us to establish links between observable 
quantities and the parameters of the SUSY model.
The superpotential yields
$F_{X_{ij}} = F_i\delta_{ij}$ and $\langle x_{ij} \rangle = 0$ for $i=j$. For 
simplicity, we shall further assume that all VEVs are real and $F_i = A M_P$ 
for all $i$, thus reducing $A$ to a single number. 
After SUSY and electroweak symmetry breaking, Eq. (5) then reduces to 
\begin{eqnarray}
 {\cal L}_{eff} = - A N^iN^i + A_{ij}
                \frac{v}{\sqrt{2}}\tilde{\nu}^i\tilde{n}^j -
        \frac{\bra x_{ij}\ket}{M_P}\frac{v}{\sqrt{2}}\nu^iN^j + {\rm h.c.},
\end{eqnarray}
where $N$ stands for right-chiral neutrino fields (and not the 
corresponding 
superfields). 
From Eq. (14) Dirac neutrino mass elements, as indicated already, are given by
\begin{eqnarray}
\left[m_D\right]_{ij} =  \frac{\bra x_{ij}\ket}{M_P}\frac{v}{\sqrt{2}}
\end{eqnarray}
while right-handed neutrino mass elements are given by
\begin{eqnarray}
\left[m_R\right]_{ij} = 2A\delta_{ij}.
\end{eqnarray}
We can immediately deduce the seesaw masses from the above via the relation
\begin{eqnarray}
m_\nu^s = -m_Dm_R^{-1}m_D^T.
\end{eqnarray}

In ${\cal L}_{eff}$ we could also use the term $\frac{1}{3!M_P^3}
[(X_{ik}^\dagger X_{kl}X_{lj}^\dagger + X_{ik} X_{kl}^{\dagger} X_{lj}^
{\dagger} + X_{ik}^{\dagger} X_{kl}^{\dagger} X_{lj}) N^iN^j]_D$ as well as 
its hermitian conjugate which are 
consistent with all conserved quantum numbers. After SUSY breaking this term 
yields
\begin{eqnarray}
{\cal{L}}_{{\tilde{n}}{\tilde{n}}} = 
{\frac {A^2}{M_P}} \bra x_{ij}\ket\tilde{n}^{i}\tilde{n}^j.
\end{eqnarray}
Consequently, the L-violating mass-squared terms for right-chiral sneutrinos 
\cite{sneu} become
\begin{eqnarray}
\Delta_{ij}^2 &=& - A^2\frac{\bra x_{ij}\ket}{M_P}.
\end{eqnarray}
\begin{figure}
\begin{center}

\includegraphics[]{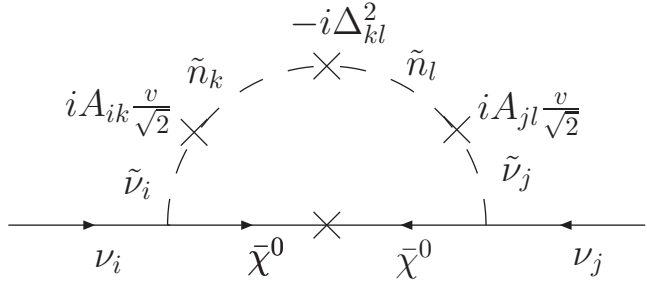}

\end{center}
\vskip -1.cm
\caption{Radiative diagram at one loop level which generates Majorana
neutrino mass.}
\end{figure}
The insertion of L-violating sneutrino masses allows the entry of  
radiative mass terms via the loop diagram shown in
Fig. 1. The expression for the loop-induced contribution is

\begin{eqnarray}
[m_\nu^r]_{ij} &=& -(A_{ik}\Delta_{kl}^2A_{jl})\frac{v^2}{2}
                \frac{g^2}{384\pi^2}\frac{1}{\tilde{m}^5}
        \nonumber \\
                &=& \frac{g^2}{384\pi^2}\frac{A^4\bra x_{ij}\ket}{M_P}
                \frac{v^2}{2}\frac{1}{\tilde{m}^5}.
\end{eqnarray}
where $\tilde{m}$ is the SUSY breaking scale in the observable sector
and is of the same order as the physical neutralino and sneutrino masses.
The origin of the parameter $A$ in this scenario has been shown in 
equation 8.
It is sufficient for us to assume $\bra x_{ij}\ket = \bra x_{ji}\ket$ and 
$A = A_i = F_{i}/M_p = F/M_P$ for all $i,j$, which 
makes each of the above matrices symmetric in $i,j$.  With
this choice and after making the transformation $\nu \rightarrow i\nu$ so as 
to change the overall sign of the neutrino mass term, the seesaw and radiative
mass matrices respectively become
\begin{eqnarray}
m_\nu^s = \frac{v^2}{4M_P}\frac{1}{F} \left(\begin{array}{ccc}
        \bra x_{12}\ket^2+\bra x_{13}\ket^2 & \bra x_{13}\ket \bra x_{23}\ket 
& \bra x_{12}\ket \bra x_{23}\ket \\
        \bra x_{13}\ket \bra x_{23}\ket & \bra x_{12}\ket^2+\bra x_{23}\ket^2 
& \bra x_{12}\ket \bra x_{13}\ket \\
        \bra x_{12}\ket \bra x_{23}\ket & \bra x_{12}\ket \bra x_{13}\ket 
& \bra x_{13}\ket^2+\bra x_{23}\ket^2
                \end{array}\right).
\end{eqnarray}
\begin{eqnarray}
m_\nu^r = - \frac{g^2}{384\pi^2}\frac{F^4}{M^5_P}\frac{v^2}{2\tilde{m}^5}
                \left(\begin{array}{ccc}
        0 & \bra x_{12}\ket & \bra x_{13}\ket \\
        \bra x_{12}\ket & 0 & \bra x_{23}\ket \\
        \bra x_{13}\ket & \bra x_{23}\ket & 0
                \end{array}\right).
\end{eqnarray}
In the above expressions we have used $A = \frac{F}{M_P}$. 
The uncertainty 
in $m_{\nu}^s$ caused by running down to the electroweak scale can be 
absorbed in $\bra x_{ij}\ket$ and $F$, since we are concerned with only the 
{\it orders of magnitude} of the latter. Also, the masses of all 
superparticles such as neutralinos and sneutrinos have been clubbed together 
as $\tilde{m}$ here. With such an approximation already made, the effect 
of renormalization group evolution is not expected to make much difference. 
Finally, with the 
effects of both nonrenormalizable interactions and lepton-number violation, 
taken into account, our most general neutrino mass matrix is

\begin{eqnarray}
m_\nu = m_\nu^s + m_\nu^r.
\end{eqnarray}

Comparing the above mass matrix with Eq. (3), we are led to the following 
equations in the notation of section 2.

\begin{eqnarray}
m_1c^2+m_2s^2 &=& \frac{v^2}{4M_P}\frac{1}{F}(\bra x_{12}\ket^2
+\bra x_{13}\ket^2) \\
\frac{1}{2}(m_1s^2+m_2c^2+m_3) &=& \frac{v^2}{4M_P}\frac{1}{F}
(\bra x_{12}\ket^2+\bra x_{23}\ket^2) \\
	&=& \frac{v^2}{4M_P}\frac{1}{F}(\bra x_{13}\ket^2+\bra x_{23}\ket^2) \\
\frac{cs}{\sqrt{2}}(-m_1+m_2) &=& - \frac{g^2}{384\pi^2}\frac{F^4}{M^5_P}
                \frac{v^2}{2\tilde{m}^5}\bra x_{12}\ket +
        \frac{v^2}{4M_P}\frac{1}{F}\bra x_{13}\ket \bra x_{23}\ket \\
\frac{cs}{\sqrt{2}}(m_1-m_2) &=& - \frac{g^2}{384\pi^2}\frac{F^4}{M^5_P}
                \frac{v^2}{2\tilde{m}^5}\bra x_{13}\ket +
        \frac{v^2}{4M_P}\frac{1}{F}\bra x_{12}\ket \bra x_{23}\ket \\
\frac{1}{2}(-m_1s^2-m_2c^2+m_3) &=& - \frac{g^2}{384\pi^2}\frac{F^4}{M^5_P}
                \frac{v^2}{2\tilde{m}^5}\bra x_{23}\ket +
        \frac{v^2}{4M_P}\frac{1}{F}\bra x_{12}\ket \bra x_{13}\ket
\end{eqnarray}
One set of consistent solutions to equations (25),(26) and (27),(28) is
$\bra x_{12}\ket = -\bra x_{13}\ket$. This reduces the above six equations 
to four which can be expressed as follows:

\begin{eqnarray}
m_1 = \frac{v^2}{2M_P}\frac{|\bra x_{12}\ket|^2}{F},
\qquad
m_2 &=& \frac{v^2}{2M_P}\frac{|\bra x_{12}\ket|^2}{F},
\qquad
m_3 = \frac{v^2}{2M_P}\frac{|\bra x_{23}\ket|^2}{F},
\qquad
        \nonumber \\
\tilde{m}^5 &=& \frac{2g^2}{384\pi^2}\frac{F^5}{M^4_P}\frac{1}
{|\bra x_{23}\ket|}
\end{eqnarray}
The last two of the above equations can be combined to eliminate 
$\bra x_{23}\ket$ and yield
\begin{eqnarray}
\tilde{m}^5 = \frac{\sqrt{2}g^2}{384\pi^2}\frac{F^5}{M^4_P}
\frac{v}{\sqrt{m_3M_PF}},
\end{eqnarray}
a form that will be used in our numerical analysis.

It is remarkable that the angle  $\theta_{12}$   
does not arise in Eq. (30).  
In the left hand sides of Eqs. (24)$-$(29) we have three independent neutrino 
mass eigenvalues and there are three independent parameters  on the 
corresponding right hand sides. The three parameters can be expressed as:
\begin{eqnarray}
\chi_1 = \frac{|\bra x_{12}\ket|}{\sqrt{F}}, \quad
\chi_2 = \frac{|\bra x_{23}\ket|}{\sqrt{F}}, \quad
\chi_3 = \frac{F^{9/2}}{\tilde{m}^5M^4}.
\end{eqnarray}
Upon using the relation $\bra x_{12}\ket = -\bra x_{13}\ket$ in 
Eqs. (24)$-$(29), 
we are left with four equations. Any three of them can be used to solve 
$\chi_1$, $\chi_2$ and $\chi_3$ in terms of $m_1$, $m_2$, $m_3$, $c$ and $s$. 
On substituting the 
values of the $\chi$'s in the fourth equation, we obtain a constraint equation 
among $m_1$, $m_2$, $m_3$, $c$ and $s$. The latter is 
automatically 
satisfied for $m_1=m_2$ irrespective of the value of $\theta_{12}$. 
Thus the near-equality of two mass eigenvalues, basically reflecting the 
smallness of the mass splitting required by the solar neutrino deficit (as 
compared to that necessitated by the atmospheric neutrino shortfall), causes 
$\theta_{12}$ to disappear from the solutions.

The above feature can perhaps be motivated by symmetries of the neutrino mass 
matrix. As has been noted in recent works \cite{Lam}, when one sets 
$\theta_{23} = \pi/4$,
$\theta_{13} = 0$, and further neglects the mass splitting $m_2 - m_1$,  
then the mass matrix becomes invariant under the successive interchange 
of the  second and third rows, and the second and third columns. This symmetry
is found to be independent of the value of $\theta_{12}$: a feature to which
the observations of the previous paragraph can be related.

\subsection{Constraint from $\mu\to e\gamma$}
Special care has been taken in our formulation to ensure a 
diagonal form for $F_{X_{ij}}$ so that FCNC processes are suppressed. 
However, a strongly constrained process like the radiative leptonic decay
 $\mu\to e\gamma$ can still 
receive one-loop 
contributions via two insertions of the L-violating sneutrino mass, as
shown in Fig. 2. 
\begin{figure}
\begin{center}

\includegraphics[]{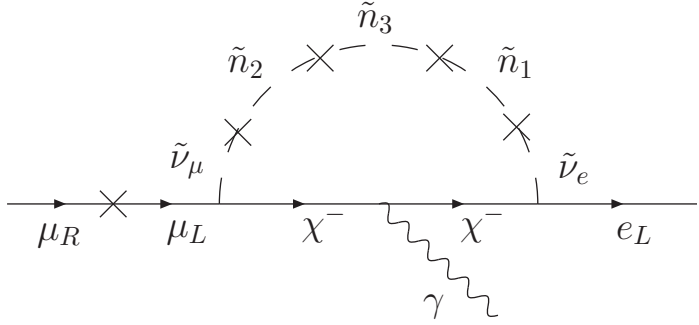}

\end{center}
\vskip -1.cm
\caption{$\mu\to e\gamma$, an FCNC process is shown at one loop level.}
\end{figure}
In order to estimate such a contribution, one can compute the corresponding
amplitude:
\begin{eqnarray}
A^\mu \sim \frac{eg^2V^{(e)}V^{(\mu)}}{16\pi^2\times30\tilde{m}}
  \left(\frac{\bra x\ket}{M_P}\right)^2 \bar{u}(p^\prime)
  \left(\frac{1+\gamma_5}{2}\right)
  (\sigma^{\mu\nu}q_\nu)u(p),
\end{eqnarray}
where $p,p^\prime$ are momenta of the incoming muon and the outgoing electron 
in Fig. 2 and $q=p^\prime - p$. Here $V^{(e)},V^{(\mu)}$ are summed mixing 
matrix elements that enter the corresponding chargino-lepton-sneutrino 
vertices. In the above equation we have used 
$A\sim M_{EW}\sim\tilde{m}$. By 
comparing this expression with, say the Standard Model amplitude for the
$\mu\to e\gamma$ transition,
we notice that there is an additional 
suppression factor of $\left(\frac{\bra x\ket}{M_P}\right)^2\sim 10^{-7}$. 
This factor is small enough to  automatically ensure a sufficiently low rate 
for the process $\mu\to e\gamma$.

\section{Different scenarios of neutrino mass hierarchy}

From the four equations (30), we notice that low-energy observables are
ultimately controlled by 
three parameters of the model, namely,  $|\bra x_{12}\ket|,|\bra x_{23}\ket|$ 
and $F$. 
In order to fix them (or the ranges they lie in), though, one needs to know
{\em the values of the neutrino masses}, along with the mixing angles. However,
while the mixing angles are experimentally known to be in certain allowed 
ranges,  
all that we can claim to know so far about the masses are the mass-squared 
differences $\Delta m^2_{12}$ and $|\Delta m^2_{23}|$, corresponding 
to the solar 
and atmospheric 
neutrino deficits respectively. Their allowed ranges of values, together with 
those of the mixing angles, can be found, 
for example,
in \cite{Strumia}. Based on these ranges, all three scenarios, namely, normal 
hierarchy, inverted hierarchy and degenerate masses \cite{Joshi}, can be 
constructed in our model. Each of these places the 
individual mass
eigenvalues within specified ranges. On using them, one can obtain the allowed 
ranges of the model parameters mentioned above. In the process, simultaneous 
use can be made of the fact that the SUSY breaking mass parameter $\tilde{m}$ 
is bounded from above if there are observable TeV-scale superparticles. 
Similarly, a lower bound on 
$\tilde{m}$ can be imposed from negative superparticle searches with present
accelerator data. 
Using the expression for $\tilde{m}$, the allowed space for 
the VEVs of the components of the gauge singlet chiral superfields can be 
further constrained. Thus one can
check whether the VEVs of the scalar and auxiliary components of
the $X_{ij}$, as required by neutrino masses, are consistent with the expected
scale of SUSY breaking and the value of the $\mu$-parameter. The self-
consistency of the entire scheme gets established in this way.

For illustration, we take the lower and 
upper bounds on $\tilde{m}$ to be about 100 GeV and 2 TeV respectively. 
The allowed mass-squared difference ranges  from oscillation data are
\begin{eqnarray}
\Delta m^2_{21} &=& (8.0\pm 0.3)\times 10^{-5} {\rm eV}^2
        \nonumber \\
|\Delta m^2_{32}| &=& (2.5\pm 0.3)\times 10^{-3} {\rm eV}^2.
\end{eqnarray}
We have taken $M_P = 2\times 10^{18}$ GeV in our numerical analysis. 
The results presented below show the minimum value of 
$\sqrt{F}$ to be above $5\times 10^{9}$ Gev, corresponding to the
lower limit on  $\tilde{m}$, which has its justification in the Large 
Electron Positron
(LEP) collider results.

\subsection{Normal hierarchy}
This scenario corresponds to 
\begin{eqnarray}
m_1\approx m_2\sim \sqrt{\Delta m^2_{21}},\qquad
m_3\sim \sqrt{|\Delta m^2_{32}|}.
\end{eqnarray}
In figures 3(a),(b) we have shown the allowed regions in the
$|\bra x_{12}\ket|-\sqrt{F}$ and $|\bra x_{23}\ket|-\sqrt{F}$
planes corresponding to the $3\sigma$ range of
$\sqrt{\Delta m^2_{21}}$ as well as of $\sqrt{|\Delta m^2_{32}|}$. 
On using the lower and upper limits of $\tilde{m}$, 
mentioned earlier, $\sqrt{F}$ is found to range between $\approx~5 
\times 10^{9}$ 
GeV and $5\times 10^{10}$ GeV. The scalar VEVs, on the other hand, are found 
to lie in the range
of $10^{11}$ $-$ $10^{12}$ GeV.  Finally, in figure 3(c)
we have plotted $\tilde{m}$ against $\sqrt{F}$ using Eq. (31) for the  
lower and upper limits  of $m_3$ at the $3\sigma$ level.
If $F$ has to be related to the SUSY breaking mass terms, it is desirable 
to have it in the high side of the allowed region shown here. Thus values of
$F$ like a few times $10^{10}$ GeV, and therefore $\tilde{m}$ somewhat on the 
higher side of
the permissible range, are therefore favored in this model, given the 
accelerator search limits
on superparticles.

The next point to note is that while $m_2$ and $m_3$ have 
specific lower as well as upper limits in the normal hierarchy scenario, $m_1$
could, in principle, go down to zero. Nevertheless, the difference between 
$m_1$ and $m_2$ being quite small, the allowed region is restricted to be 
so narrow that it can be almost called fine-tuned.
As discussed below, the situation is somewhat different in this respect in the
case of inverted hierarchy.

\begin{figure}
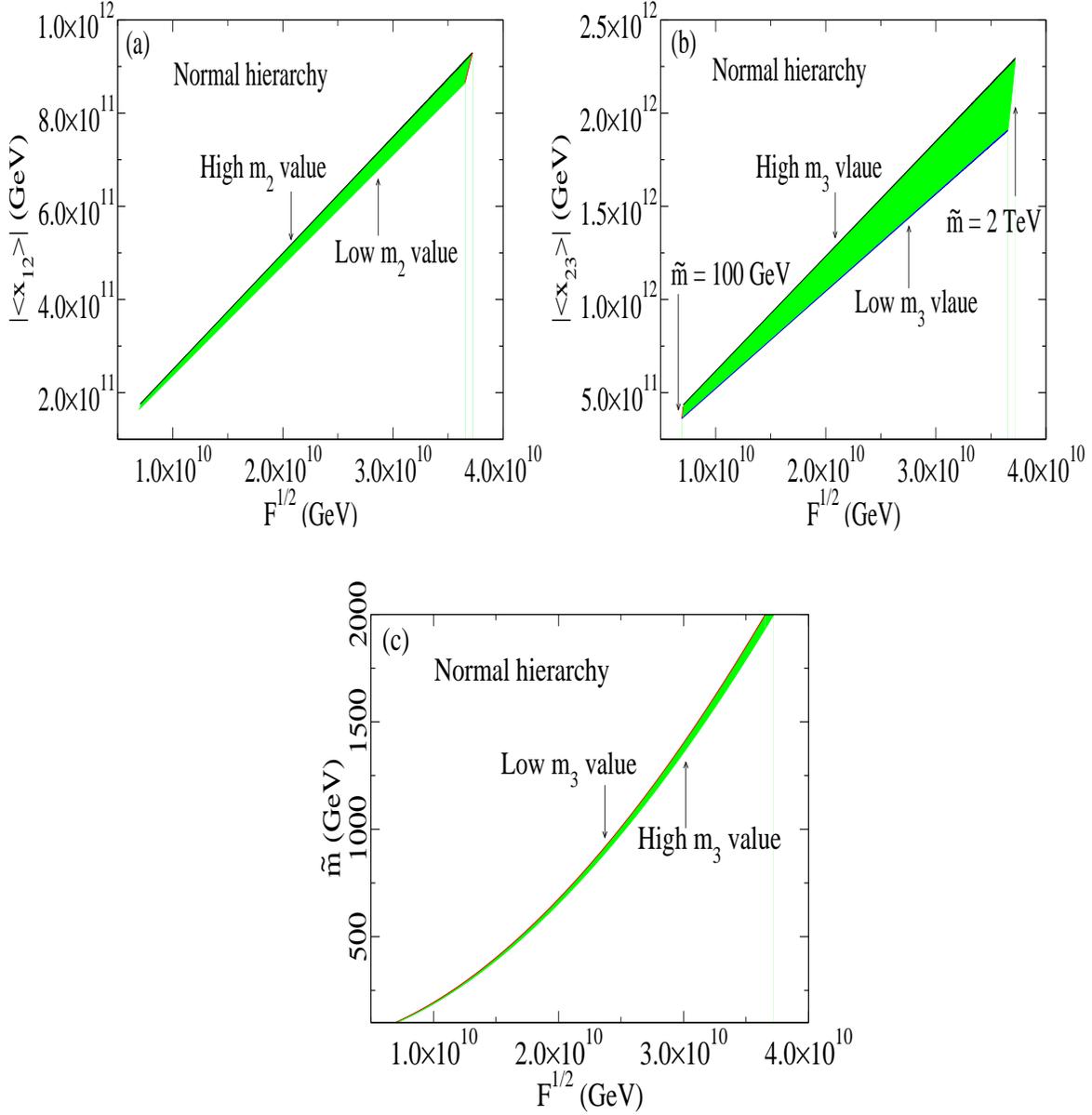

\includegraphics[height=3in,width=3in]{nh_12.eps}
\includegraphics[height=3in,width=3in]{nh_23.eps}

\begin{center}
\vskip .2cm
\includegraphics[height=3in,width=3in]{nh_m.eps}
\end{center}
\caption{Normal hierarchy: the top left (right) panel shows the allowed region
in the $|\bra x_{12}\ket| (|\bra x_{23} \ket|)-\sqrt{F}$ plane. 
The center lower panel shows the dependence of the SUSY mass scale 
$\tilde{m}$ on $\sqrt{F}$ in the allowed region of $m_3$.}
\end{figure}

\subsection{Inverted hierarchy}
Here we have
\begin{eqnarray}
m_1\approx m_2\sim \sqrt{|\Delta m^2_{32}|},\qquad
m_3\ll \sqrt{|\Delta m^2_{32}|}.
\end{eqnarray}
Notice that the relation 
$\sqrt{|\Delta m^2_{32}|} = \frac{v^2}{2M_P}\frac{|\bra x_{12}\ket|^2}{F}$
gives us the same plot as figure 3(b), with $|\bra x_{23}\ket|$ replaced by  
$|\bra x_{12}\ket|$ in the y-axis. In addition, 
$|\langle x_{23}\rangle|$ is plotted 
against $\sqrt{F}$ in figure 4(a). Since we know that
$\sqrt{|\Delta m^2_{32}|}\sim$ 0.05 eV, we have allowed a maximum of 
$m_3 = 0.01$ eV. The minimum value of $m_3$, on the other hand, could
in principle be zero. In this case, too, for each $m_3$ value there is a
limiting upper bound on $F$ coming because $\tilde{m}\leq$ 2 TeV. 
The corresponding
parameter range is represented by the shaded area in figure 4(b), where 
$\tilde{m}$ is plotted against
$\sqrt{F}$ using Eq. (31) upto a maximum value of $m_3 = 0.01$ eV 
starting from $m_3 = 0$. The interesting point to note here is that $m_3$ has
no specified lower limit in this scenario. As a result, the allowed regions 
in the
parameter space are much wider and less fine-tuned compared to the normal
hierarchy scenario. Our conclusion, therefore, is that the inverted 
hierarchy scenario
allows a larger flexibility  of high-scale parameter combinations in the scheme
adopted here.

\begin{figure}
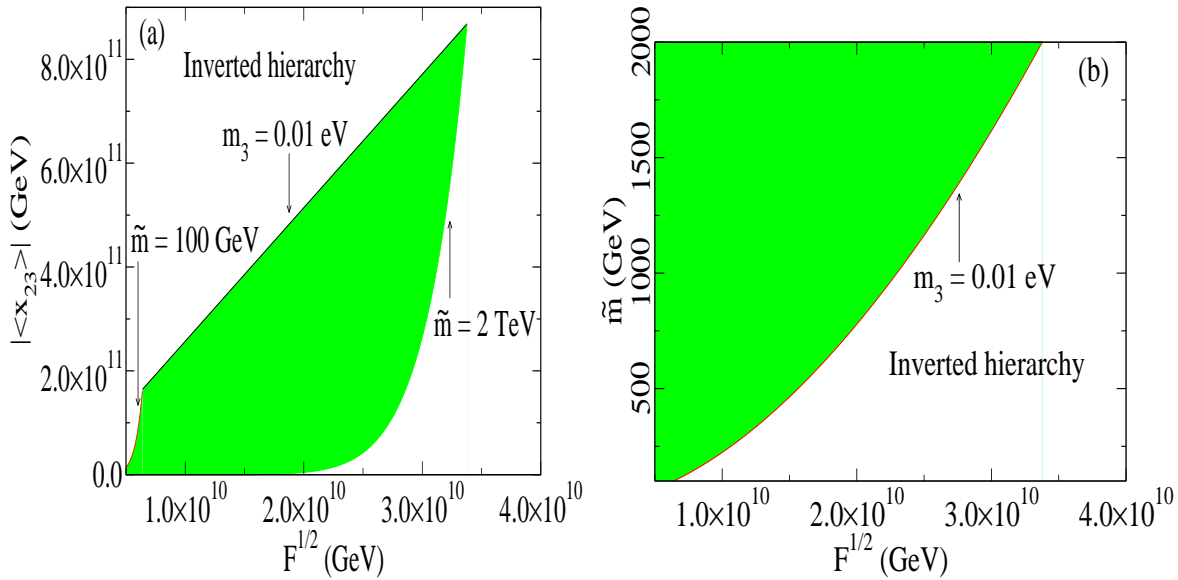

\includegraphics[height=3in,width=3in]{ih_23.eps}
\hskip .1cm
\includegraphics[height=3in,width=3in]{ih_m.eps}
\caption{Inverted hierarchy: the left panel shows the allowed region 
in the $|\bra x_{23}\ket|-\sqrt{F}$ plane. The right panel shows the 
variation of the SUSY mass scale $\tilde{m}$ against $\sqrt{F}$ in the 
allowed 
range of $m_3$.}
\end{figure}
\subsection{Degenerate masses}
This case corresponds to
\begin{eqnarray}
m_1\approx m_2\approx m_3\gg \sqrt{|\Delta m^2_{23}|},
\end{eqnarray}
leading to $|\bra x_{12}\ket|\approx |\bra x_{23}\ket|$ 
along with the requirement
the actual masses shoul be significantly 
greater than the mass-square separations. Since
$\sqrt{|\Delta m^2_{32}|}\sim$ 0.05 eV, we have to take $m_3\gg 0.05$ eV. 
On the other
hand, since these are Majorana neutrinos, there is an upper bound of about 0.3
eV on the lightest mass from
neutrinoless double beta decay as well as from cosmological constraints 
\cite{Strumia}. Therefore,
we have plotted $|\langle x_{12} \rangle|$ against $\sqrt{F}$ for 
$m_2$ ranging from 0.1 eV  to $m_2$ = 0.3 eV, showing
the allowed range as shaded area. This is shown in figure 5(a) 
with the usual constraints on $F$
coming from $\tilde{m}$. In figure 5(b), the allowed region in the 
$\tilde{m}$ $-$ $\sqrt{F}$ plane is shown with $m_3$ ranging 
between 0.1 eV and 0.3 eV.
\begin{figure}
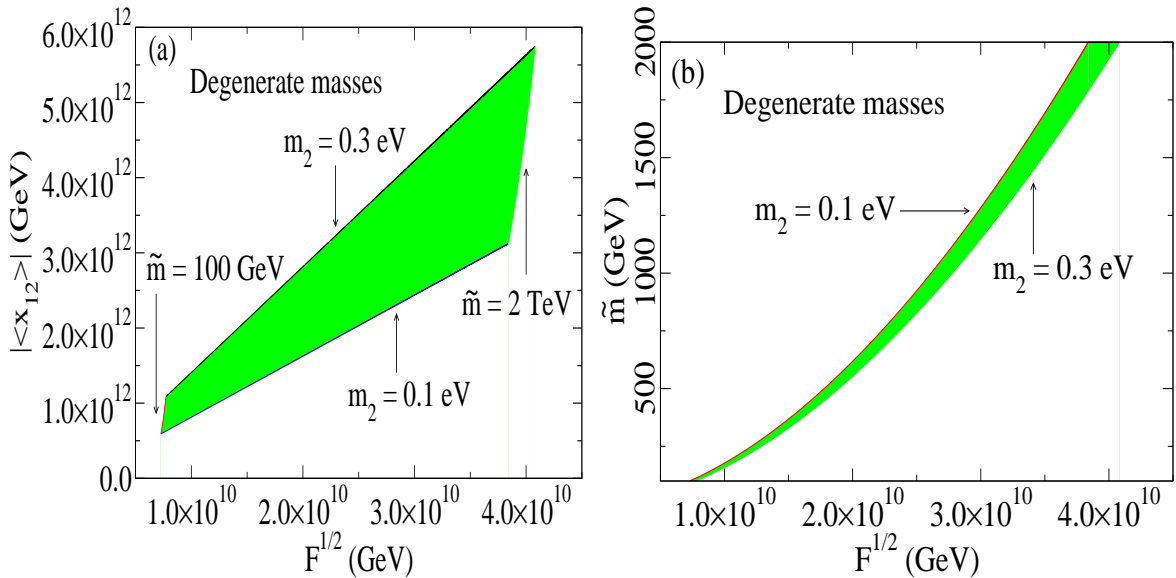

\includegraphics[height=3in,width=3in]{dm_12.eps}
\hskip .1cm
\includegraphics[height=3in,width=3in]{dm_m.eps}
\caption{Degenerate masses: the left panel shows the allowed parameter space
in the $|\bra x_{12}\ket|-\sqrt{F}$ plane. The right panel shows the
variation of the SUSY mass scale $\tilde{m}$ against $\sqrt{F}$ in the 
allowed range of $m_3$.}
\end{figure}

\subsection{Overall observations}
After analyzing these three cases, we can compare their impact on the 
parameters of our proposed model vis-a-vis the same on other similar models 
put forward in the
literature. In the model considered in \cite{Arkani2}, for example, 
there is effectively one
parameter which is $F\approx \bra x\ket^2$. This is
because just one right-chiral neutrino was considered there. That is why, 
despite the occurrence of both
seesaw and radiative masses, the former are negligibly small in magnitude
as compared to the latter. We have a more general (and natural) picture with
three right-chiral neutrino superfields. There are consequently three
unconnected parameters $F,|\bra x_{12}\ket|$ and $|\bra x_{23}\ket|$. 
These give us
more freedom enabling us to treat the seesaw and the radiative masses on
the same footing. The plots shown above for the three different neutrino 
mass cases imply that the different parameters which enter are in the expected 
range, thereby demonstrating the self-consistency of our model.

It should be noted that the SUSY breaking scale $\tilde{m}$ is related 
to $F$ by the relation $\tilde{m}\sim \frac{F}{M_P}$, and that they are not 
entirely 
independent. The results presented in figures 3$-$5 confirm that such a 
dependence is consistent with the requirement of neutrino masses and mixing. 
Some additional constraints may be required; for example, for $\tilde{m}$ 
on the higher side, one may be restricted to relatively larger values of 
$F$. However, it is impossible to be more exact in the absence of precise 
knowledge of the coupling strengths and other numerical factors in the hidden 
sector. Naively, eq. (31) is consistent with $\tilde{m}\sim \frac{F}{M_P}$ 
for $m_3$ in the range .01 $-$ .1 eV. This, in principle, restricts the 
inverted hierarchy scenario a little bit, although it is difficult to be 
very precise, for reasons already mentioned.

It is clear from the expression of Eq. (30) for $\tilde{m}$ that our model
connects very light neutrinos to TeV-scale massive particles. The latter 
include superparticles
such as neutralinos and sneutrinos as well as right-chiral neutrinos. 
Further experimental information on neutrino masses, specifically the fixation
of the hierarchy scenario, will therefore enable us to indirectly probe such 
yet undiscovered particles. 
At the same time, we have guidelines concerning the SUSY breaking sectors,
especially the non-renormalizable terms that may have other ramifications
such as explaining the worrisome $\mu$-problem.

\section{Other possibilites}

In this section we briefly comment on two other representative scenarios where
nonrenormalizable interactions may be invoked to explain the observed 
pattern in the
neutrino sector. We emphasize, however, that none of these
has any bearing on the conclusions presented in the last two sections. We 
include these remarks mainly for the sake of completeness.

In $\S3.2$ we presented a superpotential which led us to 
diagonal A-terms and the consequent suppression of FCNC effects. This requires
nonvanishing F-component 
VEVs only for the diagonal elements of the
array of superfields $X_{ij}$; the nondiagonal members of the array
should have VEVs of the scalar components only, in order to generate
off-diagonal elements of the neutrino mass matrix. One may be curious
to ask whether, for the diagonal elements $X_{ii}$, one can have nonvanishing
VEVs for both the scalar and the auxiliary components and if such be the case,
what their implications should be. 

There are some models based on Polonyi fields in which 
SUSY breaking has been achieved with a vanishing cosmological constant 
~\cite{Yanagida}. In these supergravity-inspired models, on 
integrating out some additional chiral quark fields at a scale $\Lambda$, 
an effective superpotential 
\begin{eqnarray}
W = \lambda\Lambda^2 Z 
\end{eqnarray}
is obtained at a scale below $\Lambda$. 
Here $\lambda$ is a constant ${\cal O}$(1) and $Z$ is a Polonyi 
superfield. The scalar potential constructed therefrom yields
$\langle F_Z \rangle = \lambda\Lambda^2$. At the same time, 
supergravity effects lift the 
flatness of the direction $\bra z\ket = 0$ so that both the scalar and 
auxiliary components of the chiral superfield $Z$ have nonvanishing VEVs.

Though one has in the past appealed to such models 
~\cite{Arkani1,Arkani2,Borzumati}, they do pose some difficulties in our case.
A superpotential of the above kind cannot be used
for off-diagonal elements of $X_{ij}$, since  those would generate 
large $A_{ij}$'s for $i\neq j$ and threaten to enhance FCNC rates.
Thus the superpotentials for the diagonal and off-diagonal members would
look very different in such a case, thereby raising doubts about the
legitimacy of using the components of $X_{ij}$ as fields of a similar type. 
What we have done, on 
the other hand, does not raise such questions, since the complementarity
of $\bra x_{ij}\ket$ and $F_{X_{ij}}$ is decided
essentially by the relative magnitudes of two sets of mass parameters
($\mu$, $\mu^\prime$) of the same order, where some fluctuation is quite 
natural.

Another possibility~\cite{Arkani2} lies in
considering the effective SUSY Lagrangian
\begin{eqnarray}
{\cal L}_{eff} = [XNN + LNH_u]_F + {\rm h.c.},
\end{eqnarray}
having one right-chiral neutrino to explain tiny neutrino masses. 
Eq. (39) 
can be readily generalized to include three right-chiral neutrinos, i.e. 
\begin{eqnarray}
{\cal L}_{eff} = [X_{ij}N^iN^j + \lambda_i\delta_{ij}L^iN^jH_u]_F + {\rm h.c.},
\end{eqnarray}
where there is a summation over the flavor indices $i,j$ and $\lambda_i$ 
are Yukawa couplings. The Kronecker $\delta$ in the second term is to suppress 
FCNC processes. This Lagrangian can be justified on the basis of the global 
symmetry $G_F\times G$, where $G_F = SO(3)_F$ and $G = U(1)_R\times U(1)_L$. 
For SUSY breaking, consider the hidden sector superpotential 
\begin{eqnarray}
W = \sum_{i,j=1}^3 W_{ij},
\end{eqnarray}
where
\begin{eqnarray}
W_{ij} = S_{ij}(X_{ij}\bar{X}_{ij} - \mu^2_{ij}) + \bar{X}^2_{ij}Y_{ij} 
+ X^2_{ij}\bar{Y}_{ij}.
\end{eqnarray}
The charges for various fields under $G$ in this case are:
\begin{eqnarray}
S_{ij}(2,0), \quad X_{ij}(0,2), \quad \bar{X}_{ij}(0,-2), \quad Y_{ij}(2,4),
\\ \nonumber 
\bar{Y}_{ij}(2,-4), \quad N^i(1,-1), \quad L^i(1,1), \quad H_u(0,0), 
\quad H_d(0,0).
\end{eqnarray}
The minimization of the scalar potential, arising  from this superpotential, 
yields $\bra x_{ij}\ket\neq 0$ for all $i,j$. After SUSY breaking, Dirac and 
right-handed neutrino masses are generated as
\begin{eqnarray}
[m_D]_{ij} = \frac{v}{\sqrt{2}}\lambda_i\delta_{ij}, \qquad 
[m_R]_{ij} = \bra x_{ij}\ket.
\end{eqnarray}
Hence the seesaw mass is 
\begin{eqnarray}
m_{\nu} = - m_D m_R^{-1} m_D^T.
\end{eqnarray}
For $\bra x_{ij}\ket \sim 10^{11}$ GeV and 
$\lambda_i\sim$ 0.01, we obtain neutrino masses $m_{\nu}\sim$ 0.1 eV. 
Following the analysis given in the earlier sections, we can explain the 
observed bilarge neutrino mixing as well as different hierarchies in some 
parameter space of $\bra x_{ij}\ket$. However, no connection can be made in 
this scenario between the tiny 
neutrino masses and the TeV-scale particles of MSSM. This is since only 
$\langle x\rangle$ enters the game and no $\langle F\rangle$. Moreover, 
a global symmetry of the form $U(1)_R\times U(1)_L$ does not allow the 
triple-$X$ higher order term which can generate the $\Delta L = 2$ sneutrino 
mass, cf. Eq. (9), 
in this particular model. Thus there can be no radiative contribution, at 
least not in the lowest orders. So one is unable to use here 
the full potential of such a scenario which, in our case, has meant a 
considerable widening of the parameter space through an interplay of seesaw 
and radiative effects, making our model more accommodating and natural.

\section{Summary and conclusion}

With a broken supersymmetric theory, we have considered a general scenario 
where the neutrino mass matrix is constructed
through a combination of the seesaw mechanism and radiative effects. All 
the agents behind the mechanism for this generation come from (a) a sector 
containing nonrenormalizable interactions
in the superpotential and (b) lepton-number violating terms. We have 
used an array of gauge singlet
chiral superfields $X_{ij}$ for this purpose. In terms of these, we have 
constructed a superpotential which allows both the above types of 
contributions, while
ensuring FCNC suppression. Right-chiral sneutrinos are found to have
as much of a role in the process as the corresponding neutrinos. 
It should be 
noted that the angle $\theta_{13}$ vanishes on using the
bilarge mixing matrix in equation 2. A small but nonvanishing value of
$\theta_{13}$ requires one to modify equations (24)$-$(29), although
no quatitative change in the conclusions is expected.
We have then taken in turn the neutrino mass scenarios
of normal hierarchy, inverted hierarchy and degenerate neutrinos. Using
the masses answering to each scenario, we have traced out the allowed region 
of the
parameter space of the involved high-scale physics, given in terms of the 
relevant VEVs of
the scalar and auxiliary components of the superfields $X_{ij}$. Numerically,
these are seen to allow a self-consistent region of the parameter space. While
the scenario of normal hierarchy (and, partly, that of degenerate neutrinos) 
forces us into somewhat fine-tuned zones of the parameter space, the inverted 
hierarchy picture allows a considerably larger region.
With forthcoming laboratory measurements and cosmological observations 
hopefully deciding among the above mass patterns,
connecting experimental observables to high-scale physics may become a 
realistic proposition, especially if Nature indeed proves to be supersymmetric.

{\bf Acknowledgment:} PR acknowledges the hospitality of Harish-Chandra 
Research Institute where this work was initiated. RS wishes to thank the 
Department of Theoretical Physics, Tata Institute of Fundamental Research, for
its hospitality during the progress of this work.

\end{document}